\documentclass[letterpaper]{article} %
\usepackage[draft]{aaai25}  %
\usepackage{times}  %
\usepackage{helvet}  %
\usepackage{courier}  %
\usepackage[hyphens]{url}  %
\usepackage{graphicx} %
\usepackage{natbib}  %
\usepackage{caption} %
\usepackage{bm}
\usepackage{subfig}
\usepackage{algorithm}
\usepackage{algorithmic}

\usepackage{newfloat}
\usepackage{listings}
\usepackage[acronym,shortcuts]{glossaries}
\DeclareCaptionStyle{ruled}{labelfont=normalfont,labelsep=colon,strut=off} %
\floatstyle{ruled}
\newfloat{listing}{tb}{lst}{}
\floatname{listing}{Listing}
\usepackage{amsmath,amssymb,amsfonts}

\newacronym{2D}{2D}{two-dimensional}
\newacronym{3D}{3D}{three-dimensional}
\newacronym{LiDAR}{LiDAR}{Light Detection and Ranging}
\newacronym{RI}{RI}{range image}
\newacronym{JPEG}{JPEG}{Joint Photographic Experts Group}
\newacronym{JPEG2000}{JPEG2000}{Joint Photographic Experts Group 2000}
\newacronym{INR}{INR}{implicit neural representation}
\newacronym{NNs}{NNs}{neural networks}
\newacronym{MSE}{MSE}{mean squared error}
\newacronym{BCE}{BCE}{binary cross entropy}
\newacronym{PCC}{PCC}{point cloud compression}
\newacronym{GFT}{GFT}{graph Fourier transform}
\newacronym{PCL}{PCL}{point cloud library}
\newacronym{G-PCC}{G-PCC}{geometry-based point cloud compression}
\newacronym{V-PCC}{V-PCC}{video-based point cloud compression}
\newacronym{BT}{BT}{binary-tree}
\newacronym{QT}{QT}{quad-tree}
\newacronym{ReLU}{ReLU}{Rectified Linear Unit}
\newacronym{QReLU}{QReLU}{quantum rectified linear unit}
\newacronym{NeRV}{NeRV}{Neural Representations for Videos}
\newacronym{HEIF}{HEIF}{High-Efficiency Image File Format}
\newacronym{AVIF}{AVIF}{AV1 Image File Format}
\newacronym{COIN}{COIN}{COmpression with Implicit Neural representations}
\newacronym{MPEG}{MPEG}{Moving Picture Experts Group}
\newacronym{CD}{CD}{chamfer distance}
\newacronym{R-D}{R-D}{rate-distortion}
\newacronym{BD-CD}{BD-CD}{Bj{\o}ntegaard delta chamfer distance}
\newacronym{BD-D1 PSNR}{BD-D1 PSNR}{Bj{\o}ntegaard delta D1 PSNR}
\newacronym{BD-D2 PSNR}{BD-D2 PSNR}{Bj{\o}ntegaard delta D2 PSNR}
\newacronym{DCT}{DCT}{discrete cosine transform}
\newacronym{DNN}{DNN}{deep neural network}
\newacronym{MLP}{MLP}{multi-layer perceptron}
\newacronym{NN}{NN}{neural network}
\newacronym{QML}{QML}{quantum machine learning}
\newacronym{quINR}{quINR}{quantum INR}
\newacronym{QNN}{QNN}{quantum neural network}
\newacronym{PSNR}{PSNR}{peak signal-to-noise ratio}
\newacronym{bpp}{bpp}{bit per pixel}

\title{Quantum Implicit Neural Compression}
\author {
    Takuya Fujihashi\textsuperscript{\rm 1},
    Toshiaki Koike-Akino\textsuperscript{\rm 2},
}
\affiliations {
    \textsuperscript{\rm 1}Osaka University, Suita, Osaka 565-0871, Japan\\
    \textsuperscript{\rm 2}Mitsubishi Electric Research Laboratories (MERL), Cambridge, MA 02139, USA\\
    Email: tfuji@ist.osaka-u.ac.jp, koike@merl.com
}

\begin{document}

\maketitle

\begin{abstract}
Signal compression based on implicit neural representation~(INR) is an emerging technique to represent multimedia signals with a small number of bits. 
While INR-based signal compression achieves high-quality reconstruction for relatively low-resolution signals, the accuracy of high-frequency details is significantly degraded with a small model. 
To improve the compression efficiency of INR, we introduce quantum INR~(quINR), which leverages the exponentially rich expressivity of quantum neural networks for data compression. 
Evaluations using some benchmark datasets show that the proposed quINR-based compression could improve rate-distortion performance in image compression compared with traditional codecs and classic INR-based coding methods, up to 1.2dB gain. 
\end{abstract}

\section{Background}
Representing multimedia signals (such as images and video frames) in a compact format is an important task for communicating and storing such signals. 
\Gls{INR} is an emerging memory-efficient format to compress data. 
Most \ac{INR} architectures exploit a small and simple \ac{MLP}-based \ac{NN} architecture and train the coordinate-to-value mappings using the target signals. 
For example, \ac{COIN}~\cite{dupont2021coin,dupont2022coin++} has been designed for image coding, and \ac{NeRV}~\cite{chen2021nerv} variants have considered 3D video coding. 

A key issue in such INR-based signal compression methods is the inaccurate representation of high-frequency details in a small \ac{MLP}-based \ac{NN} architecture. 
Some studies have developed sinusoidal coding~\cite{bib:mildenhall2021nerf} and activation functions~\cite{sitzmann2020implicit} to approximate high-frequency details even in a small \ac{NN} architecture. 
In this paper, we introduce a new hybrid quantum-classical \ac{INR} architecture, namely, \ac{quINR}, for signal compression.
The proposed \ac{quINR} integrates feature embedding and \ac{QNN}~\cite{farhi2018classification} for training the coordinate-to-value mapping. 
Since \ac{QNN} is a promising technique for accelerating computation and saving parameters, our \ac{quINR} may have the potential to reconstruct accurate high-frequency representations with fewer parameters.

Experiments using the \ac{RI} dataset in the KITTI \ac{LiDAR} point cloud~\cite{bib:kitti} and Kodak color image dataset~\cite{bib:Kodak} show that the proposed \ac{quINR}-based compression can provide better coding efficiency compared to the existing compression methods.  

\begin{figure*}[t]
  \begin{center}
  \subfloat[Encoding and decoding procedures using \ac{quINR}]
  {\includegraphics[width=0.95\linewidth]{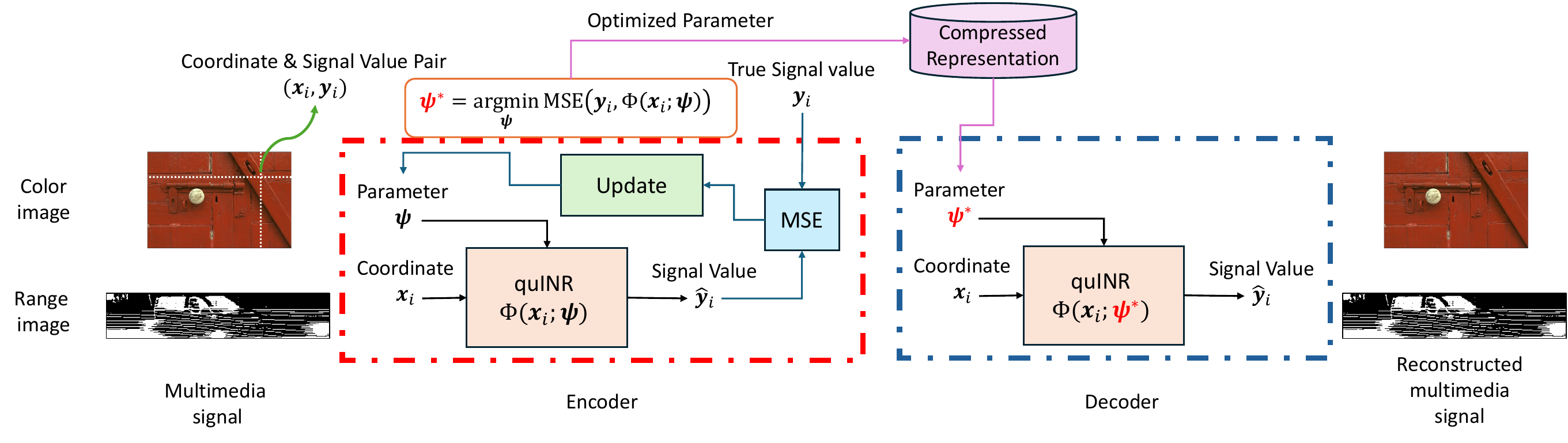}} \\
  \subfloat[Architecture of \ac{quINR}]
  {\includegraphics[width=\linewidth]{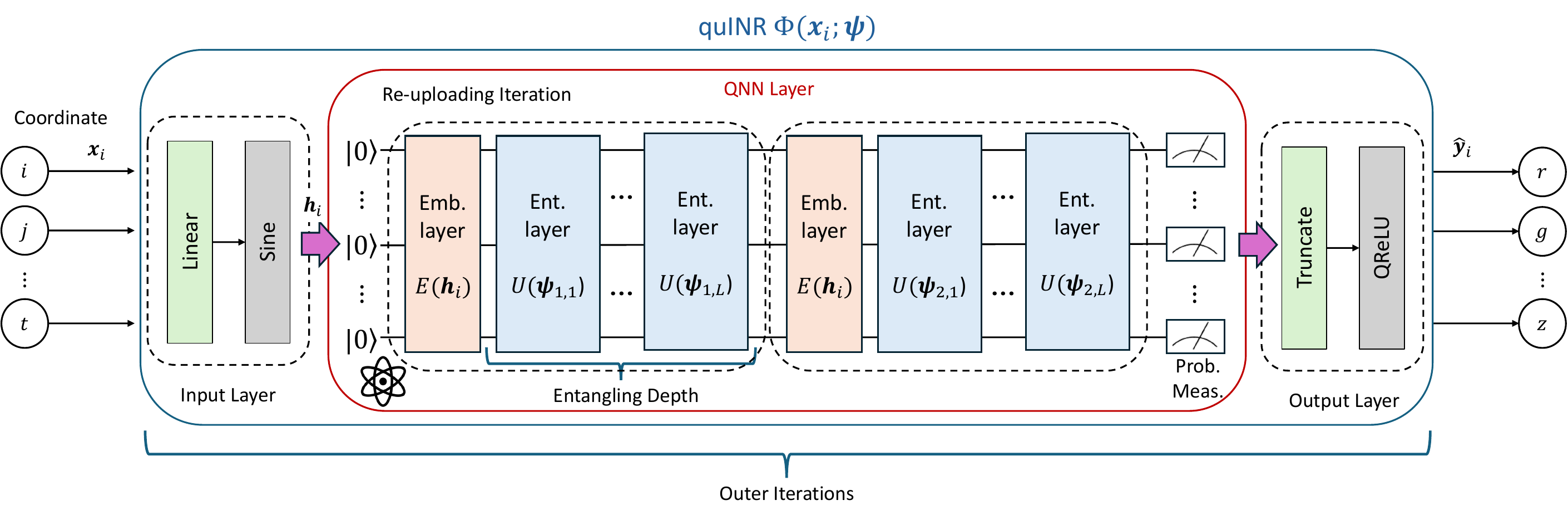}}
   \caption{Overview of the proposed scheme for data compression using hybrid quantum-classical implicit neural representation.}
   \label{fig:proposed}
  \end{center}
\end{figure*}

\section{Related Work}
\subsection{Implicit Neural Compression}
Recent studies exploit \ac{INR} architectures for data compression by overfitting a small \ac{NN} for a particular multi-dimensional sample.
The INR architecture~\cite{dupont2021coin,dupont2022coin++} takes the pixel coordinate as input to reconstruct the corresponding pixel value. 
It was extended to video coding~\cite{chen2021nerv,2024boosting} by feeding the frame index for frame generation.

\subsection{Quantum Neural Network}
\ac{QNN}~\cite{biamonte2017quantum, schuld2015introduction, farhi2018classification} is an emerging paradigm exploiting the quantum physics for neural network design, where classical data and weight values are embedded into a variational quantum circuit to control the measurement outcomes.
\ac{QNN} provides universal approximation property~\cite{perez2020data} and exponentially rich expressibity~\cite{sim2019expressibility}. 
In addition, it is analytically differentiable, enabling stochastic gradient optimization~\cite{schuld2019evaluating}.

Various frameworks were migrated into a quantum domain: 
autoencoders~\cite{romero2017quantum}; graph neural networks~\cite{zheng2021quantum}; generative adversarial networks~\cite{lloyd2018quantum, dallaire2018quantum}; contrastive learning~\cite{chen2024quantum}; diffusion models~\cite{parigi2024quantum, zhang2024generative}. 
As QNN is extremely parameter-efficient, it was applied to fine-tuning~\cite{chen2024quanta, koike2024quantum} and implicit representation~\cite{yang2022quantum, zhao2024quantum}.

\begin{figure}[t]
  \begin{center}
  \subfloat[Folded-angle embedding layer]
  {\includegraphics[trim=0 50 0 20, clip, width=0.65\linewidth]{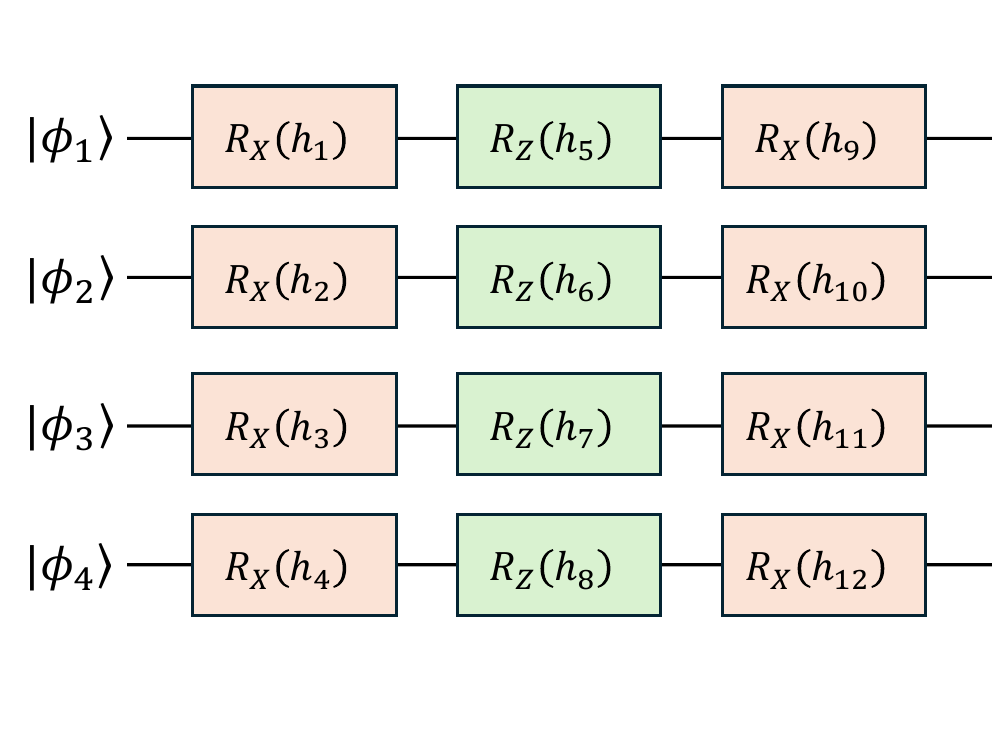}} \\
  \subfloat[Entangling layer]
  {\includegraphics[trim=0 50 0 20, clip, width=\linewidth]{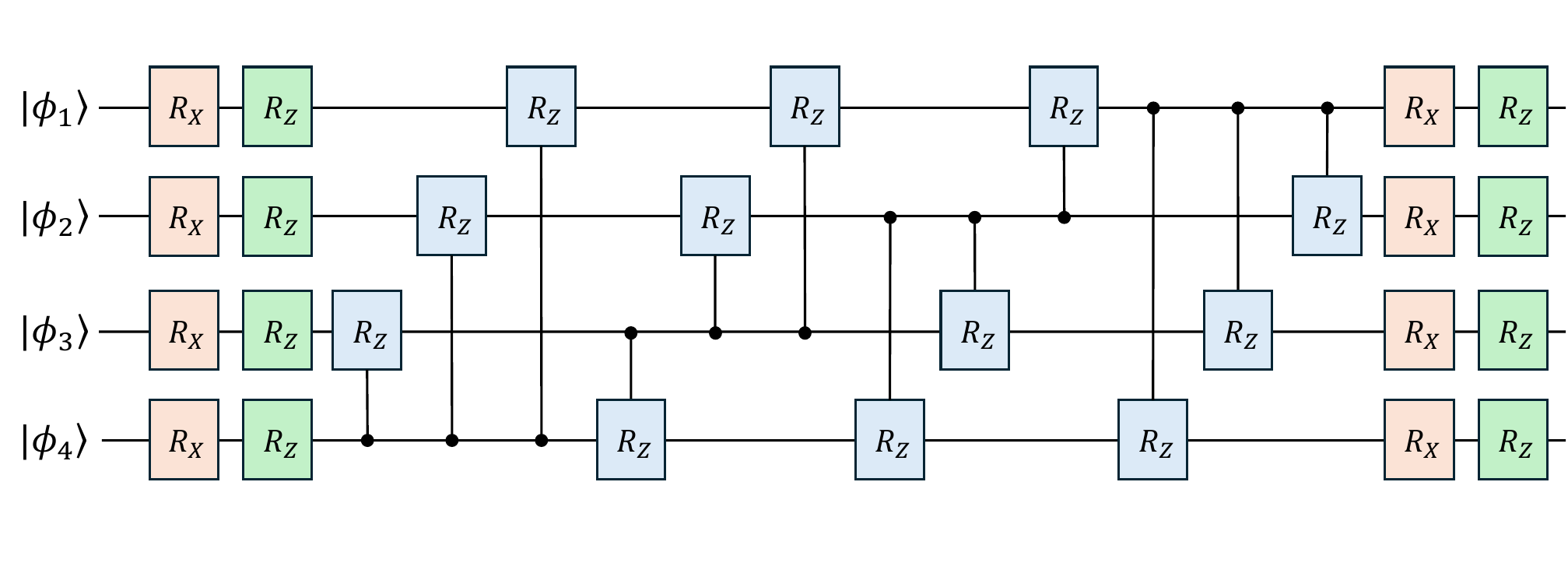}}
   \caption{Exemplar architecture of  \ac{QNN} layer.}
   \label{fig:module}
  \end{center}
\end{figure}

\section{Quantum INR for Data Compression}
\subsection{Encoding and Decoding Process}
Fig.~\ref{fig:proposed}~(a) shows the end-to-end operations of the \ac{quINR}-based encoder and decoder.
Given the target multimedia signal, we construct a dataset $\mathcal{D} = \{(\bm{x}_i, \bm{y}_i) \}$ for training the \ac{quINR} $\Phi(\bm{x}_i;\bm{\psi})$.
Here, $\bm{x}_i \in \mathbb{R}^{N_\mathrm{in}}$ is the $i$th coordinate, $\bm{y}_i \in \mathbb{R}^{N_\mathrm{out}}$ is its corresponding signal value, and $\bm{\psi}$ is the learnable parameter set.

In the encoding process, the proposed \ac{quINR} $\Phi(\bm{x}_i; \bm{\psi})$ is trained to obtain the optimized parameter set $\bm{\psi}$ to express the coordinate-to-value relationships contained in the dataset $\mathcal{D}$.
Here, we use the \ac{MSE} as the loss function to obtain the optimized parameters $\bm{\psi}$:
\begin{align}
    {\bm{\psi}}^\star = \arg\min_{\bm{\psi}} \mathsf{MSE} \big( \bm{y}_i,  \Phi(\bm{x}_i;\bm{\psi}) \big).
\end{align}
This training process is coordinate-wise, i.e., the parameters are trained to obtain a mapping from each coordinate $\bm{x}_i$ to their corresponding signal values $\bm{y}_i$.
The well-trained parameters ${\bm{\psi}}^\star$ after this encoding process are stored in storage or transmitted to the decoder as the lightweight format of the target signal.

The decoder uses the parameters ${\bm{\psi}}^\star$ for reproducing the target signal through the forward process of the \ac{quINR} $\Phi(\bm{x}_i;{\bm{\psi}}^\star)$.
The target signal $\hat{\bm{y}}_i$ is reconstructed by feeding the coordinates $\bm{x}_i$ to the \ac{quINR} architecture $\Phi(\bm{x}_i; {\bm{\psi}}^\star)$.
Likewise the encoding process, it sequentially feeds coordinates $\bm{x}_i$ to the \ac{quINR} to collect all estimated signal values $\hat{\bm{y}}_i$, which are then reshaped to the shape of the target signal as $\hat{\bm{y}}$.

\subsection{Model Architecture}
Fig.~\ref{fig:proposed}~(b) shows the proposed \ac{INR} architecture. 
The architecture takes the coordinates of the multimedia signals as inputs and generates the corresponding signal values as outputs. 
The \ac{quINR} $\Phi(\bm{x}_i;\bm{\psi})$ is a hybrid quantum-classical architecture integrating \ac{QNN} layers with a classical \ac{NN}.

The input layer consists of a linear layer with a sinusoidal activation to obtain an embedding vector $\bm{h}_i \in \mathbb{R}^{M}$ from each coordinate pair $\bm{x}_i$ as follows:
\begin{align}
    \bm{h}_i = \sin(\omega_0\bm{W}\bm{x}_i + \bm{b}),
\end{align}
where $\bm{W} \in \mathbb{R}^{M \times N_\mathrm{in}}$ and $\bm{b} \in \mathbb{R}^{M}$ are trainable parameters of the linear layer and $\omega_0=30.0$ is a constant hyperparameter.
The embedding vector $\bm{h}_i$ is then fed into the \ac{QNN} layers. 
The \ac{QNN} layers consist of embedding and entangling layers, as shown in Fig.~\ref{fig:module}. 

For embedding layer, we propose folded-angle embedding to encode an arbitrary size of embedding vector $\bm{h}_i$ into a finite number of qubits.
The conventional angle embedding has a restriction that the number of qubits must be no lower than the size of the embedding vector, while the amplitude embedding provides too small quantum space having little expressivity.
To make the QNN compact yet expressive, the folded-angle embedding uses alternating $R_X$ and $R_Z$ gates to pack more angle parameters. 
Fig.~\ref{fig:module}~(a) shows an example of $3$-folded embedding with four qubits to encode twelve variables. 

The entangling layer is based on a parameterized quantum circuit in~\cite{sim2019expressibility}. 
Specifically, the parameterized circuit sequentially carries out $R_Z$ and $R_X$ rotation gates for each qubit, two-qubit controlled Z-rotation~(CRZ) for each two-qubit combination, and finally uses Z-rotation and X-rotation. 
Here, each rotation gate is controlled based on the parameter set $\bm{\psi}$.
A few number of entangling layers are sequentially cascaded.
These embedding and entangling layers are iterated over a few layers, with a shuffled extension of the data re-uploading trick~\cite{perez2020data}.

Finally, we measure the probability value of $2^{N_q}$ quantum states.
The output layer selects the last $N_\mathrm{out}$ state with the activation function of \ac{QReLU}~\cite{qrelu}, regarded as the estimated signal value $\hat{\bm{y}}_i$. 
The above structure can be further iterated over layers to improve the capacity.

\begin{figure}[t]
  \centering
  {\includegraphics[width=\hsize]{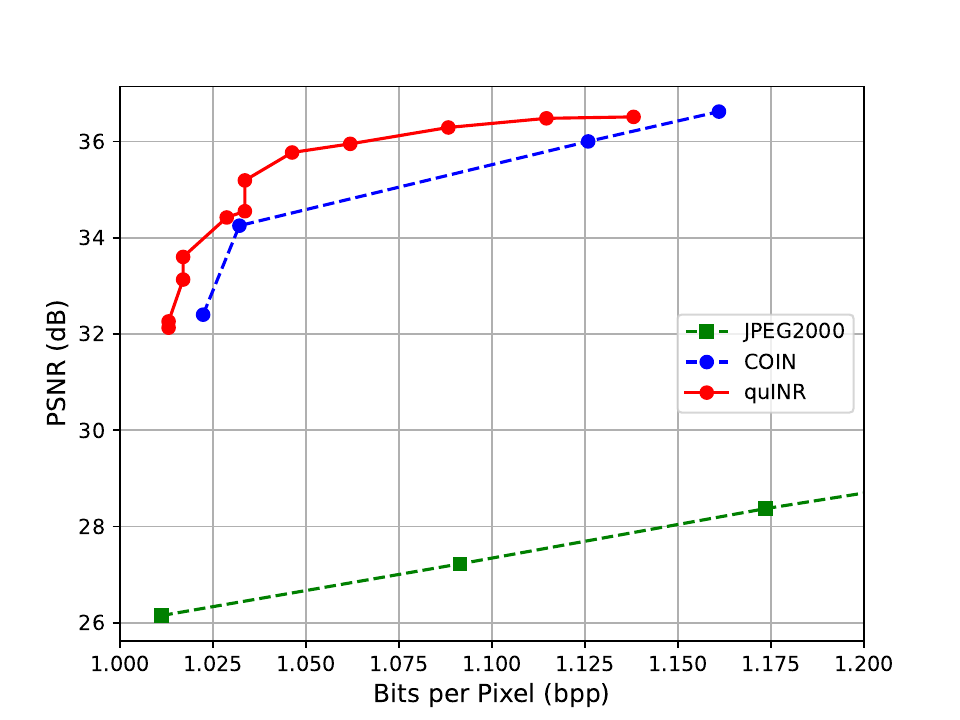}}
    \caption[]{PSNR vs.\ bpp for \ac{RI}.} 
  \label{fig:kitti}
\end{figure}

\begin{figure}[t]
  \centering
  {\includegraphics[width=\hsize]{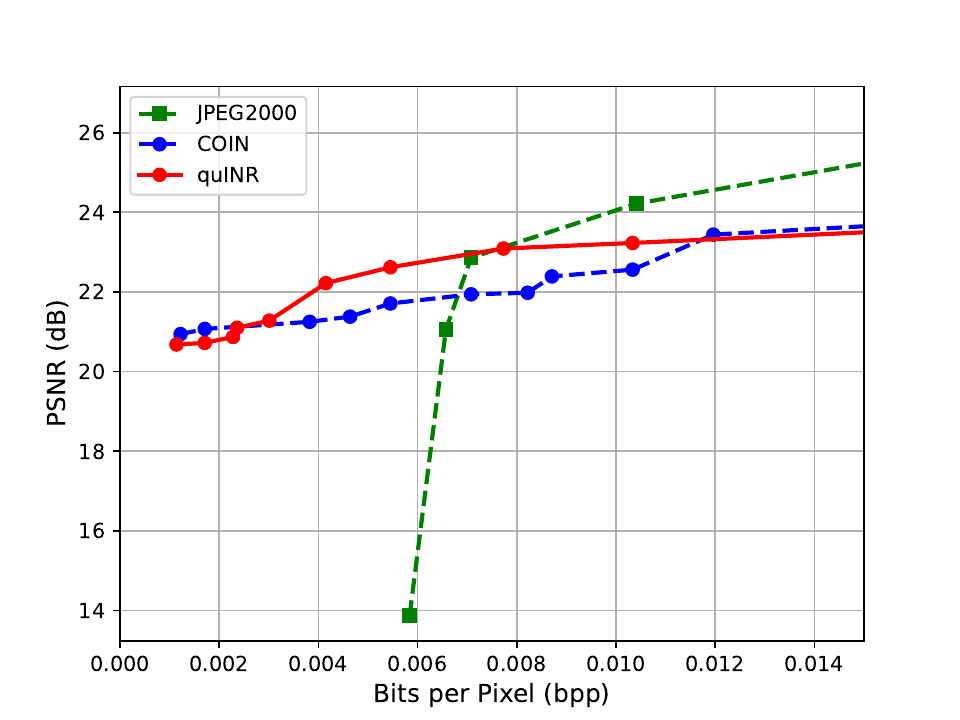}}
    \caption[]{PSNR vs.\ bpp for Kodak color image.} 
  \label{fig:kodak}
\end{figure}

\section{Experiments}
\subsection{Settings}
\subsubsection{Datasets:}
In this paper, we consider grayscale and color images to discuss the potential of the proposed \ac{quINR} architecture. 
For the grayscale image, we use \ac{LiDAR} \ac{RI}~\cite{bib:segmentation} derived from the KITTI point cloud dataset~\cite{bib:kitti}. 
\ac{RI} can be mapped from \ac{3D} Cartesian coordinate $x$-$y$-$z$ to spherical coordinate $\rho$-$\phi$-$\theta$, and then mapped to the \ac{2D} image coordinate system with the resolution of $1024 \times 64$ pixels. 
Here, each pixel value of \ac{RI} is the distance $\rho$ with floating-point precision. 
Specifically, we use \ac{LiDAR} sequence 00-00 for comparison. 
For the color image, we perform experiments on the Kodak image dataset~\cite{bib:Kodak}, which consists of 24 images of $768 \times 512$ pixels. 
We selected one image from the dataset, Kodim02.

\subsubsection{Baseline:}
We compare with baseline methods: {\ac{JPEG2000} and \ac{COIN}~\cite{dupont2021coin}. 
\ac{JPEG2000} is a typical image compression method, requiring conversion to 8-bit precision in advance for compression. 
\ac{COIN} is an \ac{INR}--based image compression baseline. The INR architecture is trained to obtain a direct mapping from the \ac{2D} pixel coordinate to the pixel value of grayscale and color images. 

\subsubsection{Implementation:}
NNs for \ac{COIN} and our proposed \ac{quINR} architectures are implemented, trained, and evaluated using PyTorch 2.0 with Python 3.9. 
The quantum circuit simulations are performed by PennyLane 0.35. 

\subsection{Performance Comparison}
Fig.~\ref{fig:kitti} shows the \ac{PSNR} performance for \ac{RI} as a function of \ac{bpp}. 
Here, we vary hyperparameters such as embedding size $M$ to show the Pareto frontier curves for each baselines. 
The results show that the proposed \ac{quINR} achieves better image quality than other baselines. 
It suggests that the proposed \ac{quINR} may have the potential to compress multimedia signals.

Fig.~\ref{fig:kodak} shows the \ac{PSNR} performance for the Kodak color image as a function of \ac{bpp}. 
For this case, \ac{JPEG2000} offers much better performance than \ac{RI} case as the target signal is a natural image.
Nevertheless, the proposed \ac{quINR} architecture can be better than the other baselines in low to medium compression regimes with up to 1.2dB gain.

\section{Conclusion}
This paper highlights the potential of quantum techniques in advancing multimedia signal compression. 
The proposed quINR architecture demonstrated good PSNR performance, particularly in compressing LiDAR \ac{RI}, leveraging the expressive power of \ac{QNN}. 
Nevertheless, its rate-distortion performance for color image compression was limited, indicating the need for further improvements, e.g., with quantum network architecture search (NAS) and distillation. 

\bibliography{aaai25}

\end{document}